\documentclass[amsmath,amssymb,aps,prd,floatfix,twocolumn]{revtex4-2}

\usepackage{amssymb}
\usepackage{epsfig,graphicx}
\usepackage{graphicx}
\usepackage{amsmath}
\usepackage{color}
\usepackage{ulem}
\usepackage{hyperref}
\hypersetup{colorlinks=true,linkcolor=blue,filecolor=magenta,urlcolor=cyan,citecolor=orange}

\usepackage{amssymb,amsmath,multirow,dcolumn,bm,float,graphicx,epstopdf}
\usepackage{calrsfs}  
\usepackage{epsfig}
\usepackage{soul}
\usepackage{color}
\usepackage{natbib}
\makeatletter

\usepackage{amssymb}
\usepackage{graphicx,color}
\usepackage[usenames,dvipsnames]{xcolor}
\usepackage{subfigure}

\colorlet{darkgreen}{green!50!black}
\colorlet{brightyellow}{yellow!75!red}%
\colorlet{orange}{red!50!yellow}
\colorlet{darkblue}{blue!60!black}
\colorlet{darkred}{red!80!black}

\renewcommand\sout{\bgroup \color{red} \ULdepth=-.5ex \ULset}

\usepackage{soul}

\usepackage{mathtools}
\usepackage{mathrsfs}
\usepackage{bm}
\usepackage{relsize}
\usepackage{indentfirst}

\usepackage{hyperref}

\def\Xint#1{\mathchoice
   {\XXint\displaystyle\textstyle{#1}}%
   {\XXint\textstyle\scriptstyle{#1}}%
   {\XXint\scriptstyle\scriptscriptstyle{#1}}%
   {\XXint\scriptscriptstyle\scriptscriptstyle{#1}}%
   \!\int}
\def\XXint#1#2#3{{\setbox0=\hbox{$#1{#2#3}{\int}$}
     \vcenter{\hbox{$#2#3$}}\kern-.5\wd0}}

\def\dashint{\Xint-}

\newcommand{\nc}{\newcommand}
\nc{\lamnr}{\lambda_{nr}}
\nc{\lamr}{\lambda_{r}}
\nc{\lamk}{\lambda_{k}}
\nc{\gp}{g({\bp})}
\nc{\gpp}{g({\bp'})}
\nc{\gpz}{g({\bp''})}
\nc{\gps}{g^{*}({\bp})}
\nc{\gpzs}{g^{*}({\bp''})}
\nc{\gpps}{g^{*}({\bp'})}
\nc{\bxi}{{\bf \xi}}
\nc{\bp}{{\bf p}}
\nc{\bpp}{{\bf p'}}
\nc{\bpz}{{\bf p''}}
\nc{\bk}{{\bf k}}
\nc{\bkp}{{\bf k'}}
\nc{\bkz}{{\bf k''}}
\nc{\bPi}{{\bf \Pi}}
\nc{\bera}{\langle}
\nc{\ket}{\rangle}
\nc{\bq}{{\bf q}}
\nc{\bqp}{{\bf q'}}
\nc{\tpi}{\tilde{\pi}}
\nc{\bpi}{\boldsymbol \pi}
\nc{\btpi}{\tilde{\boldsymbol \pi}}
\nc{\andre}[1]{\textcolor{red}{#1}}

\newcommand{\feynslash}[1]{/\hspace*{-2mm} #1}

\begin{document}
\title{Dynamical mass generation in Minkowski space at QCD scale}

\author{Dyana C. Duarte} 
\email{dyana@ita.br}
\affiliation{Instituto Tecnol\'ogico de
Aeron\'autica, 12.228-900 S\~ao Jos\'e dos Campos, SP,
Brazil.}
\author{Tobias Frederico} 
\email{tobias@ita.br}
\affiliation{Instituto Tecnol\'ogico de
Aeron\'autica, 12.228-900 S\~ao Jos\'e dos Campos, SP,
Brazil.}
\author{ Wayne de Paula } 
\email{wayne@ita.br}
\affiliation{Instituto Tecnol\'ogico de
Aeron\'autica, 12.228-900 S\~ao Jos\'e dos Campos, SP,
Brazil.}
\author{Emanuel Ydrefors}
\email{ydrefors@kth.se}
\affiliation{Institute of Modern Physics, Chinese Academy of Sciences, Lanzhou 730000, China}
\affiliation{Instituto Tecnol\'ogico de
Aeron\'autica, 12.228-900 S\~ao Jos\'e dos Campos, SP,
Brazil.}
\date{\today}
\begin{abstract}  
We undertake the challenging  task to reveal the properties of  dressed light-quarks in the space- and time-like  regions. For that aim, we solved the Dyson–Schwinger equation (DSE) in Minkowski space for the quark propagator in a QCD inspired model, focusing on the realization of dynamical chiral symmetry breaking (DCSB) in the large coupling regime.
The DSE is considered in the quenched approximation  within the rainbow-ladder truncation with a massive gluon and a Pauli-Villars term, which is used to tune the infrared (IR) physics of the model.
 The solution of the DSE in  Minkowski space is performed by resorting to the integral representation of the quark self-energy and propagator, which leads to a coupled set of closed self-consistent equations for the spectral densities, taking into account finite on-mass-shell renormalization.
 The parameters of the model are chosen such that the  gluon mass scale is consistent with recent lattice QCD (LQCD) calculations,  the Pauli-Villars mass is lowered down to about 1~GeV to concentrate strength in the infrared momentum region, and the coupling constant and renormalized mass are tuned to reproduce LQCD results for the quark mass function in the Landau gauge. 
 Future application  to
 study the pion consistently with DCSB within the Bethe-Salpeter framework with self-energies in Minkowski space is also delineated.

\end{abstract}
%

\maketitle


\section{ Introduction}
It is a challenge to establish how dynamical chiral symmetry breaking (DCSB)~\cite{Cloet:2013jya,Eichmann:2016yit,Roberts:2021nhw} manifests in Minkowski space for the light quarks starting with the Dyson-Schwinger framework. The dressing  builds the constituent quark and gluon degrees of freedom associated with a non-trivial propagator, as has  been obtained in lattice QCD calculations (see e.g. ~\cite{Oliveira:2018lln,Li:2019hyv}). This understanding is essential to provide  insights  on the partonic structure of the dressed quark, which migrates from a massive constituent particle at infrared (IR) scales to the partonic description in the ultraviolet (UV) region, as the building blocks of the Goldstone bosons associated with DCSB, like the pion and kaon, as well as the nucleon
in continuous formulations of Quantum Chromodynamics (QCD) (see e.g.~\cite{Aguilar:2019teb,Arrington:2021biu}). In this context, one main problem is the shaping of the hadron structure by the nonperturbative dynamics of QCD, which is under  intense investigation both theoretically and experimentally. It is expected that the  study of 3D imaging will clarify the hadron content, which is among the motivations for the proposal of future facilities as the Electron-Ion-Collider~\cite{AbdulKhalek:2021gbh}.
  
The enhancement of the QCD  interaction strength  at long-distances is associated with  color confinement and DCSB. It builds at the same time the  dressed quark and gluon degrees of freedom, as well as the effective interaction among them. The quark dressing  is the theory basis to justify the widely used constituent quark picture (see e.g.~the review of the quark model in~\cite{ParticleDataGroup:2020ssz}).
Extensively explored are models formulated on the light-front (LF) built with  constituent degrees of freedom of partonic nature and applied to study the hadron structure and mass spectrum~\cite{Brodsky:1997de,Bakker:2013cea}. In particular,  this conceptual framework permits to formulate confining LF  Hamiltonian models for consituent quarks and gluons,  where the corresponding hadron eigenstates are found
 by diagonalization using the basis light-front quantization (BLFQ) method~\cite{Vary:2009gt}. Contemporary applications  of this approach  to light mesons~\cite{Lan:2019vui,Adhikari:2021jrh} and the nucleon~\cite{Xu:2021wwj} explore their partonic structure.

The nonperturbative complexity of the QCD dynamics  appears  in the LF wave function, which is expanded in an infinite sum of Fock components, each of them with well-defined  number of particles and an associated probability amplitude, and these components are dynamically coupled by the interaction~\cite{Bakker:2013cea}. 
In principle, such a physically motivated representation of the hadron state and dynamics in Minkowski space is confronted by the Haag’s theorem~\cite{Haag:1955ev,Streater:1989vi}, which implies that  the formulation of the interacting and free quantum field theories can only be done on inequivalent Hilbert-space representations of the field algebra.  Therefore, at the fundamental level Haag's theorem could turn questionable the construction of the Fock-space basis from the free LF QCD Hamiltonian to describe interacting hadron states.  Recently, this issue was overcome in~\cite{Polyzou:2021qpr}  by proving the equivalence of the scattering theory formulated in instant and light-front quantum field theories, based  on a two-Hilbert space representation, without requiring the existence of a free dynamics on the Hilbert space. The  one-particle basis is built from the interacting theory, which also defines the asymptotic states. Regardless that quarks and gluons are confined and cannot be asymptotic states of QCD,  this work may suggest the dressed degrees of freedom as the constituents of the hadron LF wave function. Indeed in~\cite{Ji:2020ect}, it is proposed to build the LF probability amplitudes for $n$-partons directly from the matrix elements between the vacuum and the hadron state of correlators built  with field-operators in different positions on the light-front connected by gauge links. For example, in practice the pion LF valence wave function  was obtained from the Bethe-Salpeter amplitude projection onto the null-plane hypersurface computed in an interacting model with DCSB, where the quarks are dressed~\cite{Roberts:2012sv,Cui:2021mom}.

The main motivation for our work is  to model dynamically the dressing of the light quarks in Minkowski space, which in continuum representations of QCD are  the constituents of  light hadrons, like the pion and nucleon. We model the kernel of the Dyson-Schwinger equation (DSE) for the quark propagator to incorporate its enhancement in the IR region tuned to the relevant scales of QCD, as the gluon mass~\cite{Dudal:2013yva} and quark-gluon vertex~\cite{Rojas:2013tza,Oliveira:2018ukh,Oliveira:2020yac,Oliveira:2018fkj}, and set the parameters taking  into account the lattice QCD results for the quark mass function~\cite{Oliveira:2018lln}. 
We explore the rainbow ladder formulation of the DSE with a massive gluon-exchange in different covariant gauges including the Pauli-Villars regularization, which is tuned in a way to strengthen the kernel  in the IR region at the QCD scale. The method adopted to solve the DSE in Minkowski space is based on the K\"all\'en-Lehmann representation of the propagator and the Nakanishi integral representation (NIR)~\cite{Nakanishi} of the self-energies, which has been established as a viable tool to solve the problem.

As a matter of fact, the NIR has been already applied with success to the  solution of  the fermionic Bethe-Salpeter (BS) equation in Minkowski space since the pioneering work in Ref.~\cite{Carbonell:2010zw} and then further developed in~\cite{dePaula:2016oct,dePaula:2017ikc}, where the end-point singularities were explicitly treated. These works consolidate the technique to permit the application of NIR  to solve the pion BS equation, where it was chosen an interacting kernel  tuned to the QCD scale with constituent quarks and gluons,  and a dressed quark-gluon vertex~ \cite{dePaula:2020qna}.  It explored the momentum distributions and  3D image of the
LF valence wave function obtained by 
 projecting the BS amplitude onto the  null-plane hypersurface, also finding a probability of about 70\% for this Fock-component of the pion  state. Following it, the pion electromagnetic form factor was studied in~\cite{Ydrefors:2021dwa}, showing a nice reproduction of the experimental data, and  a fresh study with this model investigates the
 the pion parton distribution function~\cite{dePaula:2022pcb}. We should mention that other recent methods are being developed to obtain the valence wave function from the Bethe-Salpeter amplitude via  contour deformations~\cite{Eichmann:2021vnj}.

One of the first attempts to study  the dynamical mass generation of the fermion propagator in Minkowski space was made by Bicudo~\cite{Bicudo:2003fd}. The mass gap equation for spontaneous chiral symmetry breaking was solved by using an analytical approach, and   approximate solutions were obtained for the quark masses in a Yukawa model. The implications of analyticity to the solution of the DSEs  in Minkowski space was explored in~\cite{Sauli:2004bx}.
Later on,  Ref.~\cite{Sauli:2006ba}  considered a gauge theory in a quenched approximation with the massive gauge boson transverse mode, where the effective coupling was regulated by a  Pauli-Villars. According to this work, in the limit where $M/\Lambda\ll 1$ ($\Lambda$ is the Pauli-Villars mass) the analytical structure of the exact propagator is given by the Lehmann representation with one real pole. They also found that when this ratio increase, the adequate form of the exact propagator is composed by the two pole ans\"atze plus the generalized integral representation.  

The solutions to DSE  in Minkowski space using the spectral representation, for the fermion pro\-pa\-ga\-tor within quenched QED and on-shell re\-nor\-ma\-li\-za\-tion conditions in the Landau gauge was investigated in Refs.~\cite{Jia:2017niz,Sauli:2002tk}.
They obtained the analytic solutions 
using a renormalizable version of the gauge technique anzatz for the fermion-photon vertex. 

The fermion DSE for a QED-like theory was also investigated in Ref.~\cite{Frederico:2019noo} within the rainbow-ladder truncation and Pauli-Villars regularization, using methods based on the analytic continuation of the Euclidean DSE into the complex momentum plane towards the time-like region. In the first approach, properly called ``Un-Wick rotation'', the energy component of the Euclidean four-momentum is rotated in the complex plane towards the Minkowski metric. In the second approach, the Euclidean space-like four-momentum  is rotated towards the time-like one through the transformations $k\to e^{-i\delta}k, \, dk\to e^{-i\delta}dk$. When the angle $\delta = \pi/2$ these transformations retrieves the DSE in Minkowski space. The results for the model described in this work shows an excellent agreement when compared to those methods in the weak coupling regime in the Feynman gauge, for both time and space-like momentum regions~\cite{Jia:2019kbj}.

The complex analytic structure of the quark propagator in Minkowski space 
taking into account the possible existence of poles and branch cuts at time-like momenta, 
was studied in~\cite{Solis:2019fzm}, where they solve nonperturbatively
  the DSE in the strong interaction regime, using the spectral representation of the propagator and self-energy. The derived  model-independent self-consistent integral equations for the spectral functions were renormalized by  momentum-subtraction scheme.  This approach was then applied to solve the DSE in a schematic model for the quark-gluon scattering kernel.
 
In a recent publication~\cite{Mezrag:2020iuo} the coupled system of fermion and photon gap-equations in Minkowski space were formulated within the Nakanishi integral representation method. 
They derive a   coupled system of self-consistent integral equations that  allows to determine the three K\"all\'en–Lehman weights for the dressed fermion and photon propagators, and they provide a  consistency check by taking the first iteration of these equations.

The rest of this work is organized as follows. In Sect.~\ref{sec:DSEMS}, we briefly present the formulation of the coupled set of self-consistent integral equations for the spectral weights associated with the quark propagator and self-energy for covariant gauges obtained from the DSE with a massive gluon and a Pauli-Villars term. The condition for the finite on-mass-shell renormalization and the relation between the spectral densities of the self-energy and propagator are supplied in the Appendix~\ref{app1}, which are essential to close the coupled set of self-consistent equations. In Sect.~\ref{sec:ResultsLG}, we present our quantitative results, which exhibit DCSB  in the Landau gauge. We first set the model parameters at the QCD scale, for gluon and Pauli-Villars mass in order to be consistent with values suggested by LQCD calculations and inferred from studies of the quark-gluon vertex, to provide the enhancement of the kernel below 1~GeV, and reproduce to some extend the LQCD quark mass function in the space-like region. We show results for the spectral densities, the quark mass function and wave function renormalization in the time-like region.
In Sect.~\ref{sec:finalremarks} we give the final remarks on our exploratory study of DCSB and the quark dressing in Minkowski space.

\section{ DSE in Minkowski space}
\label{sec:DSEMS}

In what follows, we revise briefly the derivation of the self-consistent equation for the spectral densities of the quark propagator and self-energies in the rainbow-ladder truncation and including the Pauli-Villars term.  These set of integral equations were also used in~\cite{Frederico:2019noo,Jia:2019kbj}. This section is accompanied by the Appendix~\ref{app1}, where we present the relations between the spectral densities of the quark self-energy and propagator, as well as the on-mass-shell renormalization condition. The model adopted for the DSE for the quark propagator is written as:
\begin{equation}
\begin{aligned}
S_{q}^{-1}&(k)=\feynslash k-m_{B} \\ &+ig^{2}\hspace{-0.1cm}\int\frac{d^{4}q}{(2\pi)^{4}}\Gamma_{\mu}(q,k)S_{q}(k-q)\gamma_{\nu}D^{\mu\nu}(q)\, ,\label{SDmodel}
\end{aligned}
\end{equation}
where $m_B$ is the quark bare mass, $g$ is the coupling constant, $\Gamma^{\mu}(q,k)$ is the dressed quark-gauge-boson vertex ($\Gamma^{\mu}(q,k) = \gamma^{\mu}$ in the rainbow ladder truncation), and $D^{\mu\nu}\left(q\right)$ is the propagator of the dressed gauge boson in a covariant gauge, which is given by~\cite{Zuber1985} 
%
\begin{equation}
D^{\mu\nu}\left(q\right)=\frac{1}{q^{2}-m_g^{2}+\imath\epsilon}\left[g^{\mu\nu}-\frac{(1-\xi)q^{\mu}q^{\nu}}{q^{2}-\xi m_g^{2}+\imath\epsilon}\right]\, ,
\end{equation}
where we have introduce an effective gluon mass $m_g$, as suggested by lattice QCD (LQCD)  calculations~\cite{Dudal:2013yva}.
The quark propagator can be written as
\begin{equation}
S_{q}\left(k\right) = \Big[\feynslash k A\left(k^{2}\right) - B\left(k^{2}\right)+i\epsilon\Big]^ {-1}\, ,\label{Sf1}
\end{equation}
and the vector and scalar self-energies are given by the  NIR, respectively as: 
\begin{eqnarray}
A\left(k^{2}\right) & = & 1+  \int_{0}^{\infty}ds\frac{\rho_{A}\left(s\right)}{k^{2}-s+i\epsilon}\, ,\label{Af}\\
B\left(k^{2}\right) & = & m_B + \int_{0}^{\infty}ds\frac{\rho_{B}\left(s\right)}{k^{2}-s+i\epsilon}\, .\label{Bf}
\end{eqnarray}
The self-energy normalization conditions are such that $A(\pm\infty) = 1$ and $B(\pm\infty) = m_B$. In our model the finite bare mass is identified with the current quark mass. 

The pole of Eq.~\eqref{Sf1} is the renormalized mass $\overline m_0$. The on-mass-shell renormalization condition relates the bare mass $m_B$ with $\overline{m}_0$ through:
\begin{equation}
A^2(\overline{m}_0^2) - B^2(\overline{m}_0^2) = 0\,,
\end{equation}
which is written in terms of $\overline{m}_{0}$, $m_B$ and the spectral weights, $\rho_A$ and $\rho_B$, and it is explicitly given by Eq.~\eqref{mbare}.

The quark propagator is then written in terms of  the vector, $S_v(k^ 2)$, and scalar, $S_s(k^ 2)$, components in the form:
\begin{align}
 S_{q}(k) &= \feynslash k S_v(k^2) + S_s(k^2) \nonumber\\
 &=  R\frac{\feynslash k+\overline{m}_{0}}{k^{2}-\overline{m}_{0}^{2}+i\epsilon}+\feynslash k\int_{0}^{\infty}ds\frac{\rho_{v}\left(s\right)}{k^{2}-s+i\epsilon} \nonumber\\
 &+ \int_{0}^{\infty}ds\frac{\rho_{s}\left(s\right)}{k^{2}-s+i\epsilon}\, ,\label{srenf}
\end{align}
where $R$ is the residue at the renormalized mass pole,  which is one of the inputs in the model.

Introducing Eq.~\eqref{Sf1} in \eqref{SDmodel} after standard Dirac algebra it is possible to write the vector and scalar self-energies in a covariant gauge $\xi$ as
\begin{align}
&k^{2}A\left(k^{2}\right)  =i\alpha\int\frac{d^{4}q}{4\pi^{3}}\Biggl\{\frac{R}{\left(k-q\right)^{2}-\overline{m}_{0}^{2}+\imath\epsilon}\nonumber \\
 & +\int_{0}^{\infty}ds\frac{\rho_{v}\left(s\right)}{\left(k-q\right)^{2}-s+\imath\epsilon}\Biggl[\frac{-2k^{2}+2\left(k\cdot q\right)}{q^{2}-m_g^{2}+\imath\epsilon}\nonumber \\
 & -\frac{(1-\xi)\left(2\left(k\cdot q\right)^{2}-k^{2}q^{2}-q^{2}\left(k\cdot q\right)\right)}{\left(q^{2}-m_g^{2}+\imath\epsilon\right)\left(q^{2}-\xi m_g^{2}+\imath\epsilon\right)}\Biggl]\Biggl\}\nonumber\\
&- [m_g\to \Lambda] \,,\label{k2A2}
\end{align}
and 
\begin{align}
B\left(k^{2}\right) & =-i\alpha\int\frac{d^{4}q}{4\pi^3}\Biggl\{\frac{R\,\overline{m}_{0}}{\left(k-q\right)^{2}-\overline{m}_{0}^{2}+\imath\epsilon}\nonumber \\
 & +\int_{0}^{\infty}ds\frac{\rho_{s}\left(s\right)}{\left(k-q\right)^{2}-s+\imath\epsilon}\nonumber \\
 & \times\frac{1}{q^{2}-m_g^{2}+\imath\epsilon}\left[4-\frac{q^{2}(1-\xi)}{q^{2}-\xi m_g^{2}+\imath\epsilon}\right]\Bigg\}\nonumber\\
 &- [m_g\to \Lambda]\,,\label{Bk2}
\end{align}
where we have made use of Eq.~\eqref{srenf} with the K\"all\'en-Lehman spectral representation  of the scalar and vector components of the quark propagator.
In these equations $\alpha = g^2/(4\pi)$, with $g$ being the quark-gluon coupling constant.
We introduce a Pauli-Villars regulator by subtracting a term with $m_g$  replaced by $\Lambda$ in the integrand  of Eqs.~\eqref{k2A2}  and \eqref{Bk2}, which also cancels the logarithmic divergences in the loop momentum integrals. In particular, in the Landau gauge, $\xi=0$, and Feynman  gauge, $\xi=1$, the contribution of the Pauli-Villars regulator can be effectively associated with the form-factor of the $\gamma^\mu$ component of quark-gluon vertex:
\begin{equation}
    \lambda_1(q^2)=\frac{\Lambda^ 2-m_g^2}{q^2-\Lambda^2+i\epsilon}\, ,\label{Lambda1}
\end{equation}
which gives us a guidance on the range of values around $\Lambda\sim 1$~GeV to be used in the calculations performed in this work, following the study from Ref.~\cite{Rojas:2013tza,Oliveira:2018ukh,Oliveira:2020yac}.

Although we will  specifically present numerical results for the Landau gauge, as the model will be parametrized to reproduce LQCD results in that gauge, the equations are derived for arbitrary covariant ones, and due to that it is useful to redefine the equations above as:
\begin{equation}
\begin{aligned}
A\left(k^{2}\right) &= A_{F}\left(k^{2}\right)+A_{\xi}\left(k^{2}\right)\, ,\\
B\left(k^{2}\right)  &=  B_{F}\left(k^{2}\right)+B_{\xi}\left(k^{2}\right)\, ,\label{ABarb}
\end{aligned}
\end{equation}
where $A_F,B_F$ correspond to Feynman gauge ($\xi = 1$) and $A_{\xi},B_{\xi}$ correspond to the  $\xi-$gauge contributions.

 Using the Feynman parametric formula it is possible to perform the momentum integrals in Eqs.~(\ref{k2A2})-(\ref{Bk2}). After that by evaluating the imaginary part of both sides of these equations, taking into account the integral representation of $A(k^2)$ and $B(k^2)$ given in Eqs.~\eqref{Af} and \eqref{Bf}, respectively, one finds that, 
\begin{align}
&\rho_{A}(\gamma)  = R\mathcal{K}^\xi_{0A}(\gamma, \overline{m}_{0}^{2},m_g^{2})\nonumber \\
 & +\int_{0}^{\infty}ds\,\mathcal{K}^\xi_{A}(\gamma,s,m_g^{2})\,\rho_{v}(s) -\left[m_g\to\Lambda\right]\, ,\label{rhoAfull}\\
 &\rho_{B}(\gamma)  = R\,\overline{m}_{0}\,\mathcal{K}^\xi_{0B}(\gamma,\overline{m}_{0}^{2},m_g^{2})\nonumber \\
 & +\int_{0}^{\infty}ds\,\mathcal{K}^\xi_{B}(\gamma,s,m_g^2)\,\rho_{s}(s) -\left[m_g\to\Lambda\right]\, ,\label{rhoBfull}
\end{align}
where  
$ \mathcal{K}^\xi_{0A(0B)}= K_{A(B)}
 +m_g^{-2}\bar K_{A(B)}^{\xi}$ is the driving term. The kernel
 is
 \begin{align}
 & \mathcal{K}^\xi_{A}(\gamma,s,m_g^{2})= K_{A}(\gamma,s,m_g^{2})\,\Theta(s-(\overline m_0+m_g)^2)
  \nonumber \\
 &+m_g^{-2}\bar K_{A}^{\xi}(\gamma,s,m_g^{2})\, \Theta(s-(\overline m_0+\sqrt{\xi}m_g)^2)\, ,
\label{KAxi} \end{align}
with the analogous formula for $\mathcal{K}^\xi_{B}$.  
We define the variable $\Delta=\gamma-m_g^{2}+s$ and with that the kernels are written as
\begin{align}
&K_{A}\left(\gamma,s,m_g^{2}\right) \hspace{-0.1cm}=\hspace{-0.1cm} -\frac{\alpha}{4\pi}\frac{\Delta}{\gamma^2}
\sqrt{\Delta^{2}-4\gamma\, s}
\nonumber\\
&\times\Theta\left[\gamma-(m_g +\sqrt{ s})^2\right]\,,\label{KernFeynA}\\
&K_{B}\left(\gamma,s,m_g^{2}\right)\hspace{-0.1cm}=\hspace{-0.1cm}-\frac{\alpha}{4\pi}\frac{4}{\gamma}\sqrt{\Delta^{2}-4\gamma s}\nonumber\\
&\times\Theta\left[\gamma-(m_g +\sqrt{ s})^2\right]\, ,\label{KernFeynB}
\end{align}
which corresponds to the  Feynman gauge kernel, $\xi=1$, and
\begin{small}
 \begin{align}
&\bar K_{A}^{\xi}(\gamma,s,m_g^2) = -\frac{\alpha}{4\pi}
\frac{\left(\gamma-s\right)^{2}-m_g^2
(\gamma+s)}{2\gamma^{2}}\nonumber\\
&\times\,\sqrt{\Delta^{2}-4\gamma s}
\,\Theta\left[\gamma-(m_g\hspace{-.1cm} +\hspace{-.1cm}\sqrt{ s})^2\right]-[m_g^{2}\to\xi m_g^{2}]
\, ,
\label{KernCsiA}\\
&\bar K_{B}^{\xi}(\gamma,s,m_g^{2}) = \frac{\alpha \,m_g^{2}}{4\pi}\frac{\sqrt{\Delta^{2}-4 \gamma s}}{\gamma}\nonumber\\&\hspace{1.8cm}\times\Theta\left[\gamma-(m_g\hspace{-.1cm} +\hspace{-.1cm}\sqrt{ s})^2\right]-[m_g^{2}\to\xi m_g^{2}]\, ,\label{KernCsiB}    
\end{align}
\end{small}

\noindent
which is nonzero in arbitrary $\xi$ gauges, except for the Feynman one.  
It is worth to mention that the Heaviside step function $\Theta$ in the kernels defines the upper limit for the integration in $s$ in the definitions of $\rho_A$ and $\rho_B$. 

Furthermore, due to the theta functions in the kernel and driving terms of Eqs.~\eqref{rhoAfull} and \eqref{rhoBfull}, we have that: 
\begin{equation}
    \rho_v(\gamma)=\rho_s(\gamma)=\rho_A(\gamma)=\rho_B(\gamma)= 0\, ,
\end{equation}
for $\gamma<(\overline m_0+\sqrt{\xi}m_g)^2$ in the case of $\Lambda>m_g$. It is important to observe that our formalism is valid for $\xi\geq0$. 

The residue of the propagator $R$ is evaluated as
\begin{align}
R^{-1}  =1&+\dashint_{0}^{\infty}\hspace{-0.2cm}ds\frac{\rho_{A}(s)}{\overline{m}_{0}^{2}-s} -2\overline{m}_{0}^{2}\dashint_{0}^{\infty}\hspace{-0.2cm}ds\frac{\rho_{A}(s)}{(\overline{m}_{0}^{2}-s)^{2}}\nonumber\\
 &+2\overline{m}_{0}\dashint_{0}^{\infty}\hspace{-0.2cm}ds\frac{\rho_{B}(s)}{(\overline{m}_{0}^{2}-s)^{2}}\,,\label{resid}
\end{align}
where the symbol $\dashint$ represents the Cauchy principal value. 
The details about the derivation of this expression, as well as, the relation between the spectral densities $\rho_v(\gamma),\,\,\rho_s(\gamma),\,\,\rho_A(\gamma)$ and $\rho_B(\gamma)$ are given in Appendix~\ref{app1}.

\section{ Results in Landau gauge} 
\label{sec:ResultsLG}
The dynamical quark mass $M(k^{2})$ and the wave function renormalization $Z(k^2)$ are defined respectively as 
\begin{align}
M(k^{2})  =\frac{B(k^{2})}{A(k^{2})}\quad\text{and}\quad Z(k^{2}) & =\frac{1}{A(k^{2})}\label{wavef}\,.
\end{align}
Within a phenomenological perspective, the parameters of the model will be tuned for a comparison with LQCD results in the Landau gauge~\cite{Oliveira:2018lln} for the quark  mass function and wave function renormalization, as expressed by the parametrizations shown in~\cite{Oliveira:2020yac}.
This allows the possibility to explore  dynamical chiral symmetry breaking in Minkowski space at the QCD scale. 
For this model one may expect that chiral symmetry is broken when the bare mass $m_B \simeq 0$ and $B(k^2)\neq 0$~\cite{Sauli:2006ba}.
It is worth to emphasize that in this approach we use the renormalized mass $\overline{m}_0$ as the input, while the bare mass $m_B$ is evaluated from Eq.~(\ref{mbare}).

\begin{table}[t]
\caption{Different sets of input parameters: renormalized mass, $\overline{m}_0$, gluon mass $m_g$, Pauli-Villars mass $\Lambda$, and $\alpha$ used in the figures, for Landau gauge. 
The parameters are adjusted in order to get the running mass close to the fit to the Lattice QCD results in Landau gauge, as shown in Fig.~\ref{runM}. The outputs quantities, the bare mass $m_B$ and residue of the propagator $R$ at the renormalized 
mass $\overline{m}_0$}
\begin{center}
\begin{tabular}{cccccc}
\hline
 Set       & $\overline{m}_0$ (GeV)& $m_g$ (GeV) & $\Lambda$ (GeV)  & $\alpha$ \\ \hline
1   & 0.42                  & 0.84               & 1.20     & 19.70        \\
2   & 0.38                  & 0.78               & 1.10     & 20.30        \\
 3   & 0.35                  & 0.60               & 1.00    & 13.25        \\
\hline
\hline
Set &  (outputs) & $m_B$ (MeV) & $R$ \\
\hline
1   &  & 9.29        & 2.22 \\
2   &  & 8.78        & 2.09 \\
3   &  & 11.92       & 2.64 \\
\hline
\end{tabular}
\end{center}
\label{parameters}
\end{table}%

 We present in  Table~\ref{parameters} three different sets of parameters for our model with results for the quark mass function comparable to the fit to LQCD calculations~\cite{Oliveira:2018ukh,Oliveira:2020yac}. The main guidance to our search of the parameters, is the gluon effective mass around 0.6~GeV~\cite{Dudal:2013yva} and the parameter $\Lambda$ around 1~GeV.  The  renormalized quark mass and the effective $\alpha$ are found to reproduce to some extend the LQCD quark mass function in the space-like region and in a way also to provide the small bare masses of about 10~MeV.

\begin{figure}[t]
\centering 
\includegraphics[width=.8\linewidth]{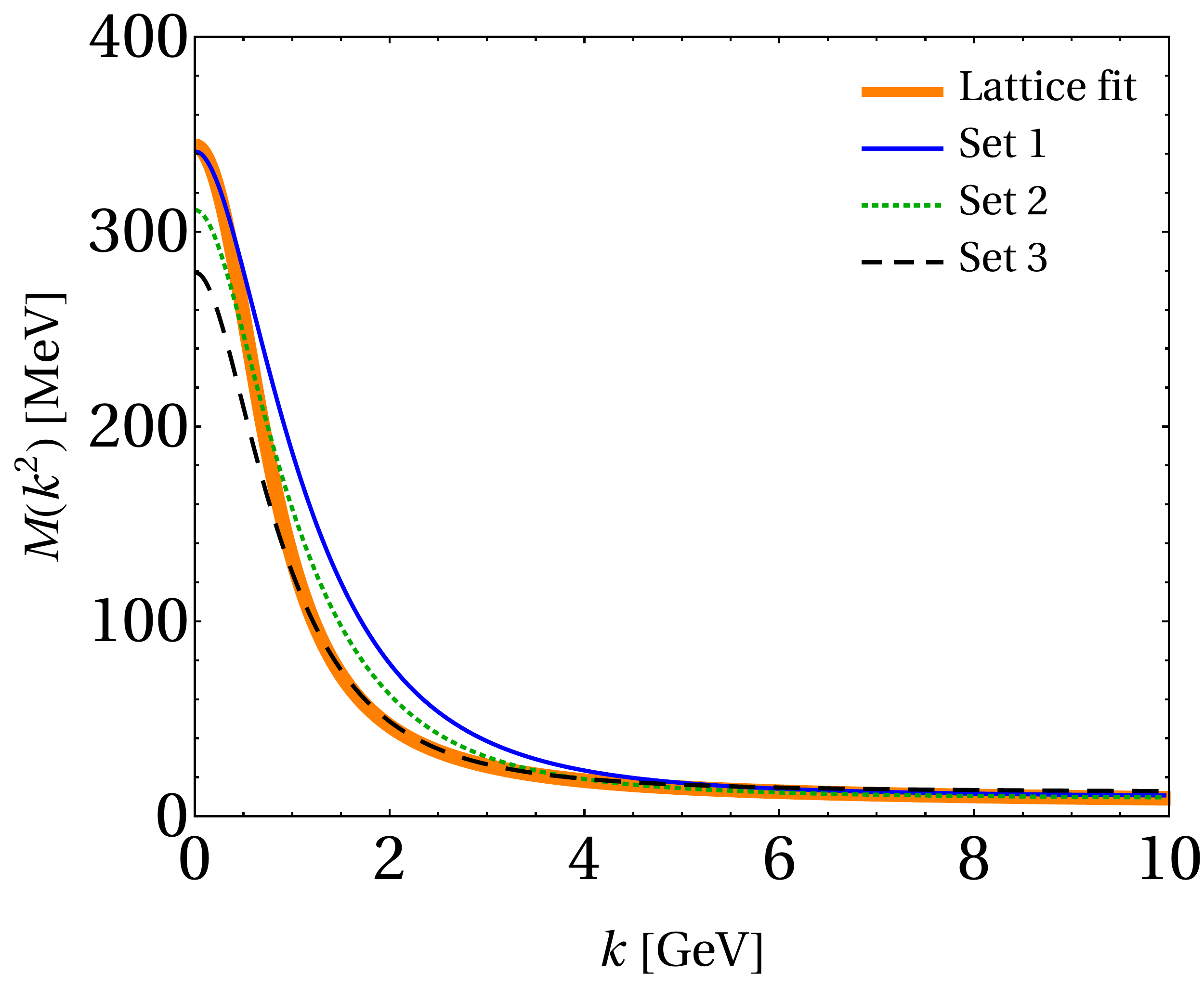}
\includegraphics[width=.8\linewidth]{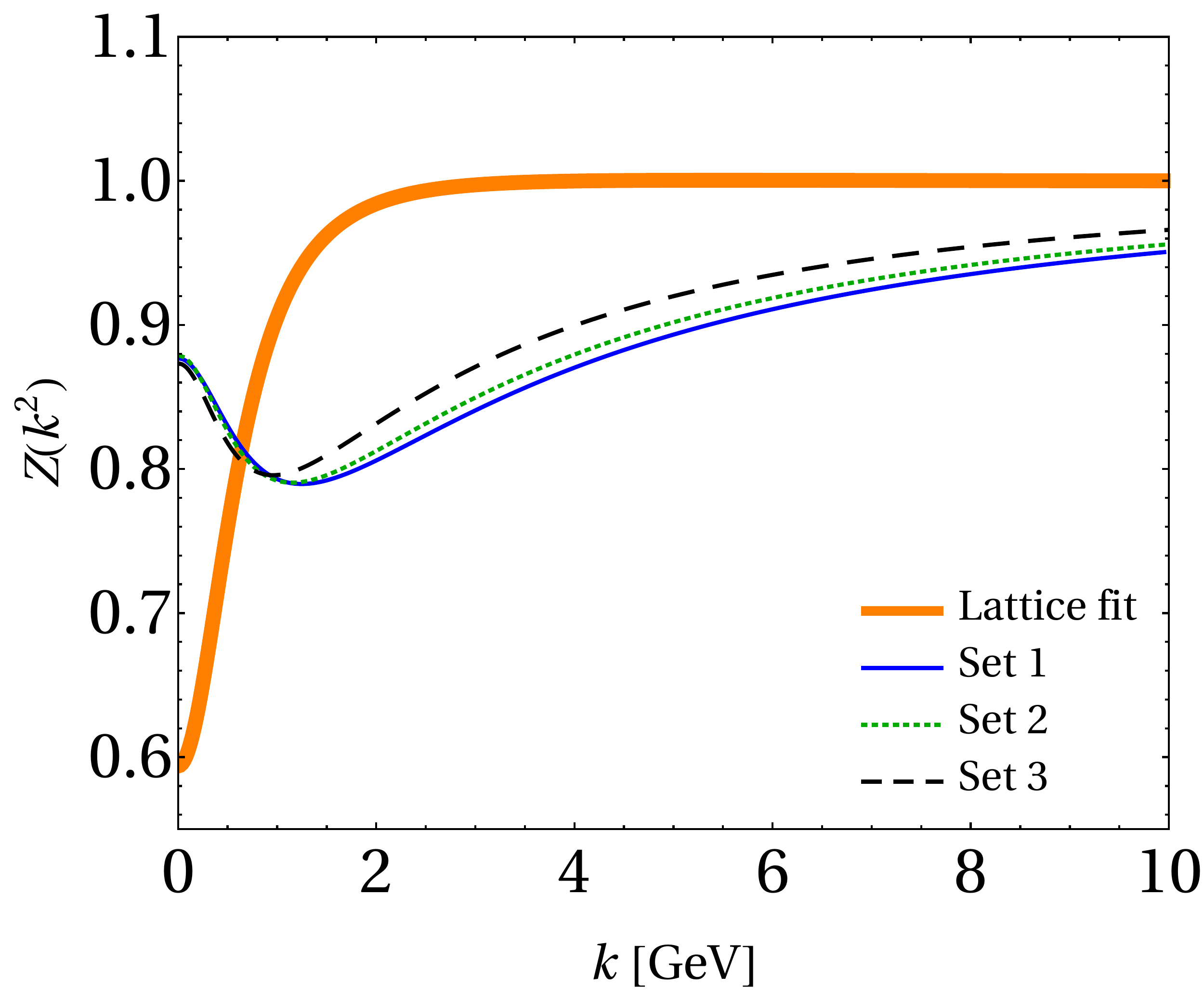}
\caption{Landau gauge results for the running mass $M(k^2)$  and quark wave function $Z(k^2)$ as functions of spacelike momentum $k$, using the sets of parameters given in Table~\ref{parameters}. Solid thick curves are the fit of LQCD calculations for the mass function and wave function renormalization given in~\cite{Oliveira:2018lln}.}
\label{runM}
\end{figure}

 \begin{figure}[t]
\centering 

\includegraphics[width=.8\linewidth]{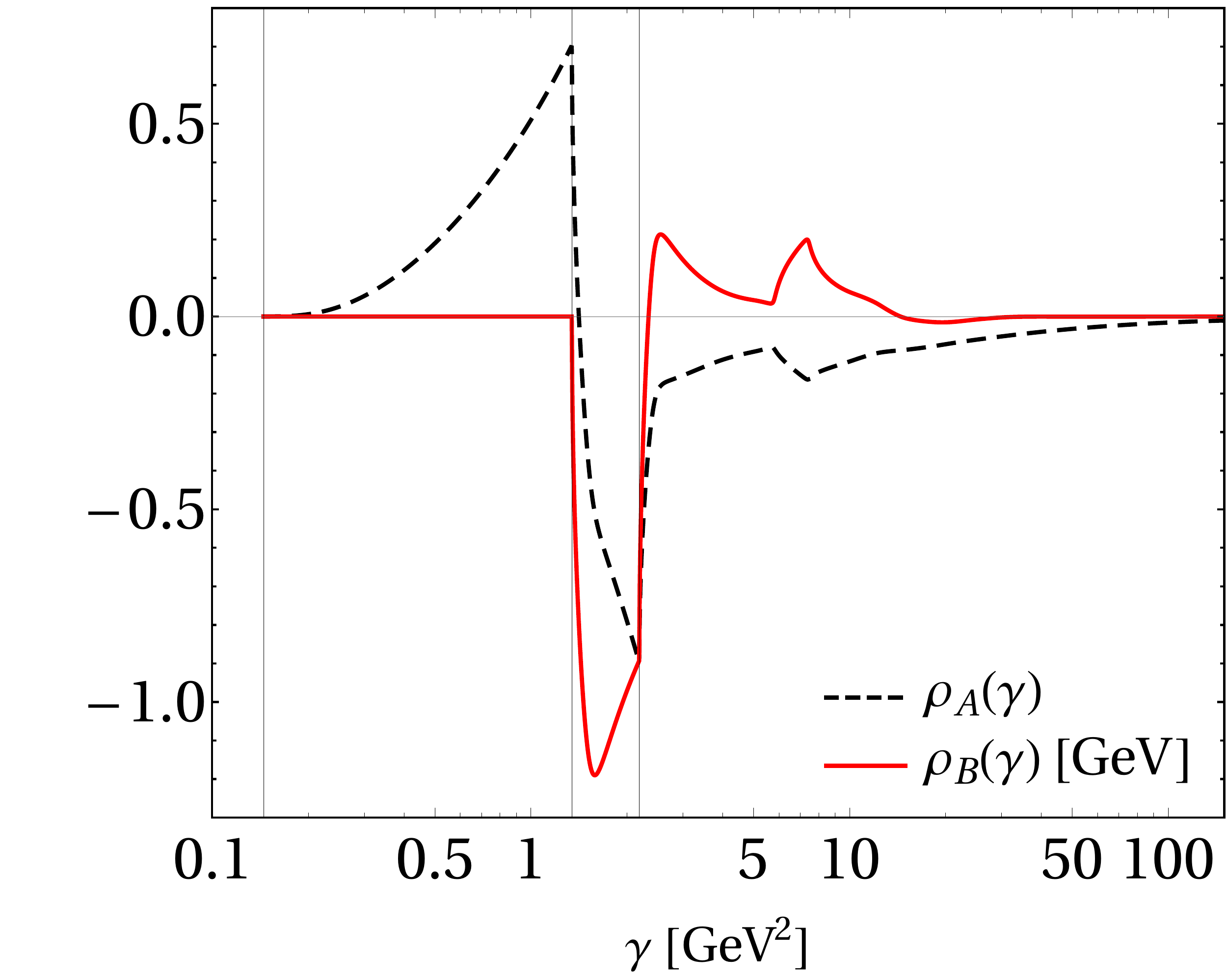}

\vspace{0.25cm}

\includegraphics[width=.8\linewidth]{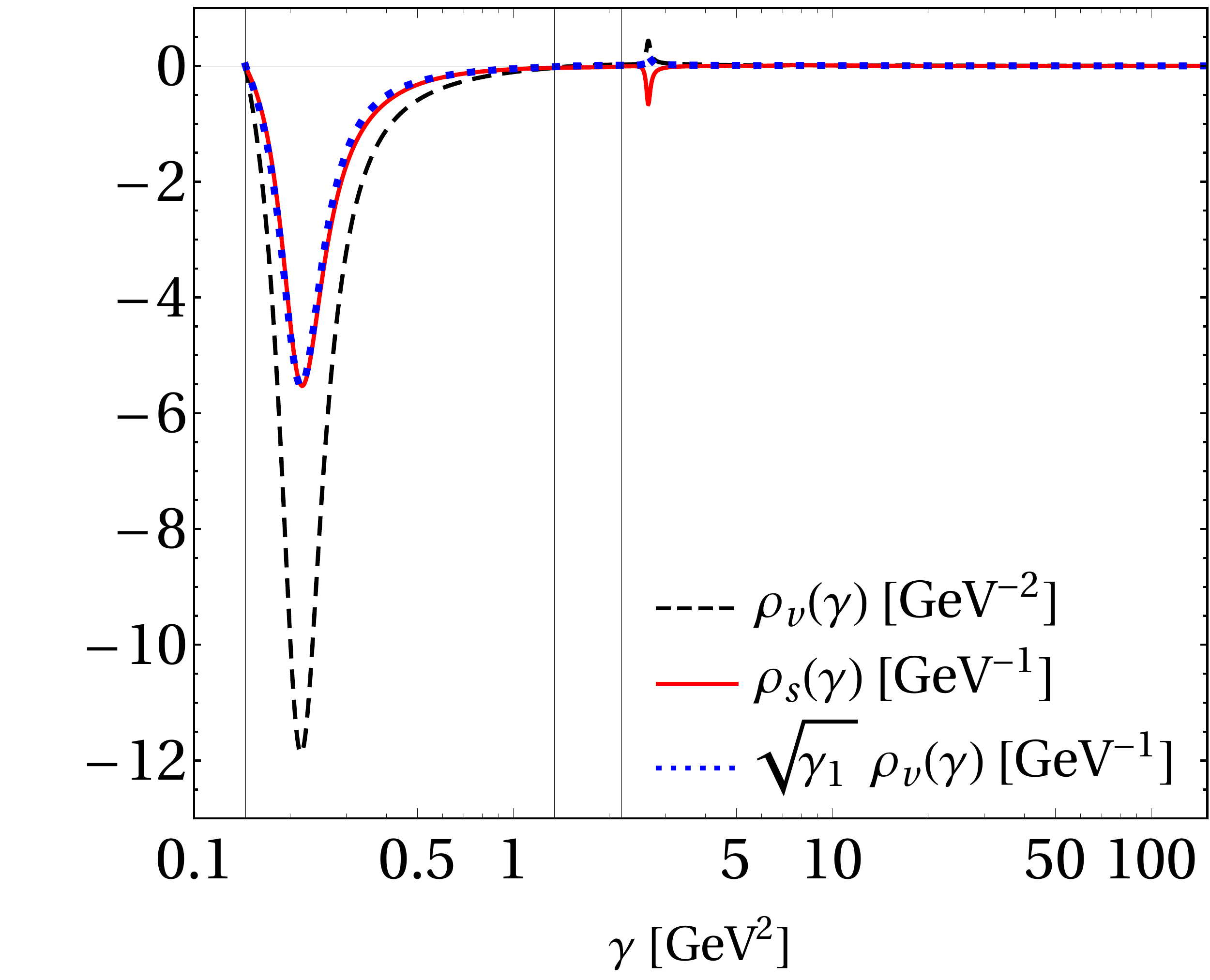}

\caption{ Spectral densities for the self-energy  (upper panel) and for the propagator (lower panel) as functions of $\gamma$, computed for set 2 from Table~\ref{parameters}.   The vertical lines are $\overline m^2_0$, $(\overline m_0+m_g)^2$ and $(\overline m_0+\Lambda)^ 2$ from the thresholds of the driving terms in Eqs.~\eqref{rhoAfull} and \eqref{rhoBfull}. ($\gamma_1=0.216$~GeV$^2$, see text).  }
\label{rho}
\end{figure}

We should mention that in Refs.~\cite{Oliveira:2018ukh,Oliveira:2020yac} it was found 
an infrared enhancement of the quark-gluon vertex by combining the DSE with Slavnov-Taylor identities and lattice simulations,
with a peak of $\alpha_s\lambda_1\sim 5.5$ for momentum around $\Lambda_{\text{QCD}}$, which can be related  to our kernel parametrization as follows.
The quark-gluon form factor in Eq.~\eqref{Lambda1} at $q^2=0$ gives for the kernel strength $\alpha\,(1-m_g^2/\Lambda^2)$, which is 10, 10.1  and 8.48 for set 1, 2 and 3, respectively,
these values are somewhat of the order to what were found in the quoted references.

The infrared physics within the model effectively parameterizes the main phenomena of the DCSB. Such a finding is consistent with the studies of Refs.~\cite{Oliveira:2018ukh,Oliveira:2020yac,Oliveira:2018fkj}, where the relevance of the infrared physics appeared through the enhancement of the longitudinal components of the quark-gluon vertex for momenta around $\Lambda_{\text{QCD}}$. In these studies the longitudinal components were constrained by the Slanov-Taylor identity in consistence with the Dyson-Schwinger equation for the quark self-energy and the corresponding LQCD results. We also compare the renormalized mass pole and residue to the fitting of the mass function with a monopole-like form~ \cite{Mello:2017mor}, which has the lowest mass pole at  0.327~GeV with residue 1.49, comparable to the present results shown in Table~\ref{parameters}.

Fig.~\ref{runM} compares our results for  the quark mass function and wave function renormalization in the space-like momentum region for the three sets of Table~\ref{parameters}  to the fit of LQCD calculations~\cite{Oliveira:2018lln}. While set 1 (thin solid curve) well reproduces the mass function in the low momentum region, in the range $0\leq k\lesssim 1$~GeV, set 3 (dashed curve) shows better agreement with lattice results for $k$ above 1~GeV. At large momentum the calculated mass function tends to the bare mass around 10~MeV.
 The strong IR effects captured by the model is still found at momentum somewhat above the ones exhibit by LQCD calculations.

 From the comparison of the wave function renormalization in the lower panel of Fig.~\ref{runM}, we see that  the model presents a pronounced  dip around 1~GeV - a qualitative property found in LQCD calculations close to $k=0$. In addition, we observe that the minimum of our solution for $Z(k^2)$ is dislocated to lower values from set 1 to 3, as both $\overline m_0$ and $\Lambda$ decreases (cf.~Table~\ref{parameters}). This suggests that effectively the IR scale accommodated within the model  is somewhat larger than $\Lambda_{\text{QCD}}$. This also explains why the model is not able to simultaneously reproduce the mass function in the IR and for $k$ around 1-2~GeV: if we privilege this last region we miss the IR behavior of LQCD results as in set 3. On the contrary, if the mass function reproduces the IR region it overestimate the mass function of above 1~GeV, as happens for set 1.

 This behavior can be understood by inspecting the kernel of the self-consistent integral equations ~\eqref{rhoAfull} and \eqref{rhoBfull}, which in the Landau gauge is non vanishing above  $\gamma =\overline m^2_0$, due to the theta functions in Eq.~\eqref{KAxi}. Ob\-vi\-ous\-ly,   the region below  $\overline m_0 \sim \Lambda_{\text{QCD}}$ is not explored by the model. That is why the minimum of wave function renormalization is shifted to larger values around $\overline m_0\lesssim k_\text{min}\lesssim \Lambda$. Indeed, by decreasing $\overline m_0$ and $\Lambda$ we observe that the minimum is driven to lower values of the space-like momentum seen in the lower panel of Fig.~\ref{runM}, as we have already pointed out before.

In Fig.~\ref{rho} we show the  results for  the self-energy and for  the propagator spectral densities obtained with parameter set 2 from Table~\ref{parameters}.
The driving terms of Eqs.~(\ref{rhoAfull}) and~(\ref{rhoBfull}) for $\xi=0$ due to the theta functions have the first threshold  at $\gamma = \overline{m}_0^2$, followed by $\gamma = (\overline{m}_0 + m_g)^2$ and $\gamma = (\overline{m}_0 + \Lambda)^2$, which are represented by the vertical lines in the figure. 
In the upper panel of the figure  one  sees that the spectral densities for the self-energy present discontinuities in their derivatives at the thresholds. 

In the lower panel of Fig.~\ref{rho}, we observe two peaks/depths on both $\rho_v$ and $\rho_s$  located at $\gamma_1 = 0.216$~GeV$^2$ and $\gamma_2 = 2.66$~GeV$^2$. Notice that $\sqrt{\gamma_1}\rho_v\approx\rho_s$, which can be checked by comparing the solid line with the dotted line. This suggests to interpret $\sqrt{\gamma_1}=0.465$~GeV as somewhat analogous to a mass pole, if you compare the spectral density with the one for the Dirac propagator of a structureless fermion, with, however, a negative  residue, which violates the positivity constraints~\cite{Zuber1985}.
Interestingly enough the solution to Eqs.~\eqref{rhoAfull} and ~\eqref{rhoBfull} pile-up strength attempting to produce a second mass pole but with an unphysical negative residue.  The quark is not an asymptotic state of QCD as it is confined and therefore such a feature of the model interaction is not at odds with the physics we want to describe. The integral of $\rho_v(\gamma)$ around the negative peak at 0.465~GeV gives -1.42, comparable to
the residue of -0.58 at the second mass pole of 0.644~GeV obtained from the fitting of the quark mass function with a monopole-like form~\cite{Mello:2017mor}.  Noteworthy that in Ref.~\cite{Jia:2017niz} the scalar weight is also negative, as well as in \cite{Mezrag:2020iuo}, which was  attributed to the choice of the renormalization scheme.

The sharp increase and subsequent peak at $\gamma\sim \gamma_1(\gamma_2)$ in $\rho_B(\gamma)$ can be traced back to the   peaks and depth  on  $\rho_s$, at $\gamma_1$ and
$\gamma_2$ ($\sqrt{\gamma_2}=1.63$~GeV). We observe peaks in $\rho_B$ at 
$\gamma\sim 3$~GeV$^2$ and
$\sim 7.37$~GeV$^2$, which are  located close to $(\sqrt{\gamma_1} + \Lambda)^2=2.4$~GeV$^2$ and $(\sqrt{\gamma_2} + \Lambda)^2=7.5$~GeV$^2$, respectively. As a conclusion,  the ``shark fin'' behavior seen in $\rho_B(\gamma)$ (upper panel)
could be  qualitatively related  with the two new ``thresholds'' at $\approx$~2.4 and 7.5~GeV$^2$. Such a result is associated with  the Pauli-Villars contribution to the kernel. A similar behavior is observed in $\rho_A(\gamma)$, but with the opposite sign. The subtraction of the Pauli-Villars term in the kernel of Eqs.~\eqref{rhoAfull} and \eqref{rhoBfull} has the physical effect of concentrating its strength in the IR region characterized by the lowest threshold  $\overline m_0^ 2< \gamma\lesssim (\overline m_0+\Lambda)^2$, corresponding to $0.38<\sqrt{\gamma}\lesssim 1.48$~GeV, which has a visible  effect in both $\rho_A$ and $\rho_B$ below $\sim 2$~GeV$^2$. Furthermore,
the equivalent quark-gluon vertex  form factor $\lambda_1$ in the Landau gauge, Eq.~\eqref{Lambda1}, has a characteristic momentum scale of $\Lambda=1.1$~GeV for our analysis of set 2 (cf. Table~\ref{parameters}).

 \begin{figure}[t]
\centering 
\includegraphics[width=.8\linewidth]{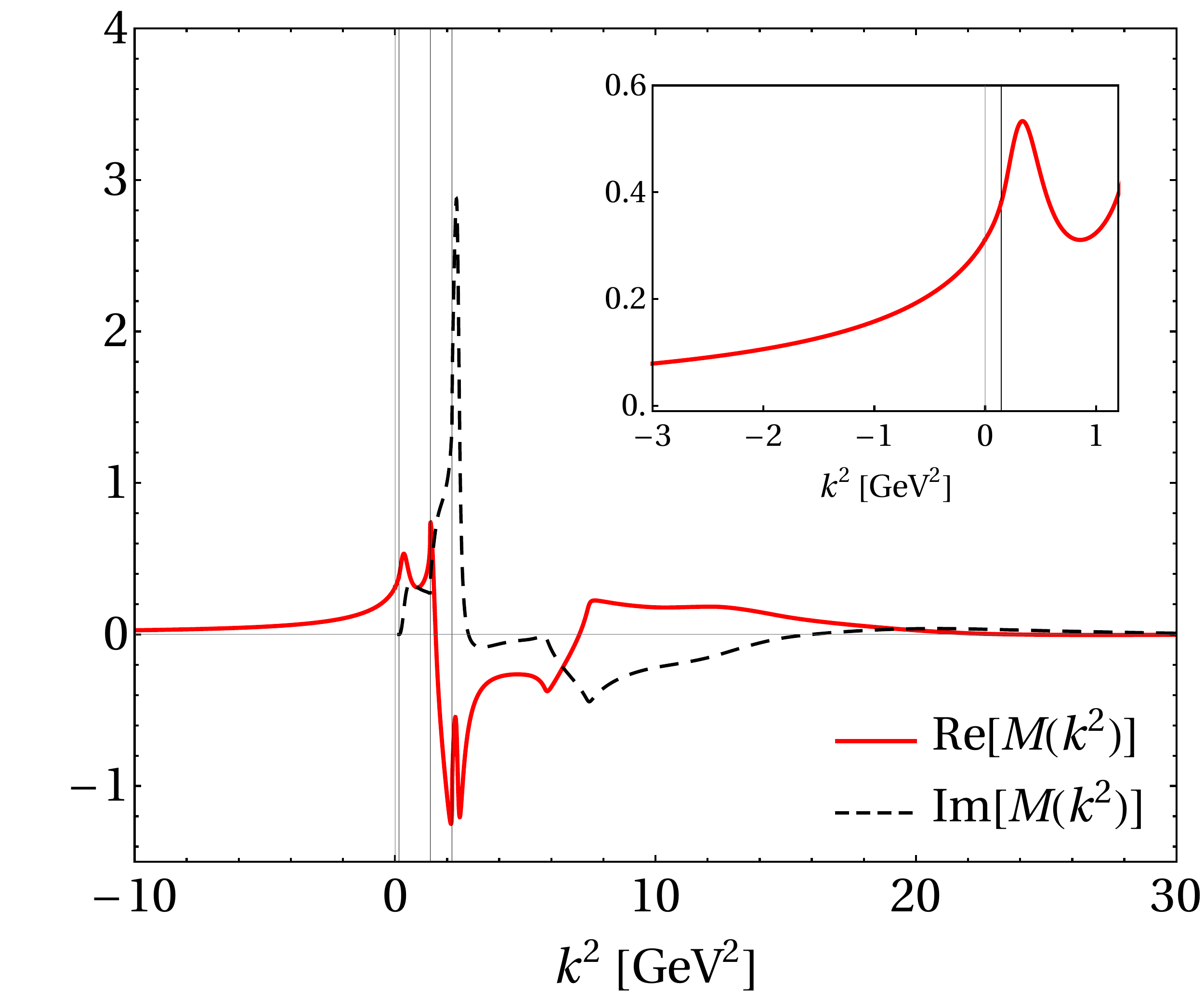}
\includegraphics[width=.8\linewidth]{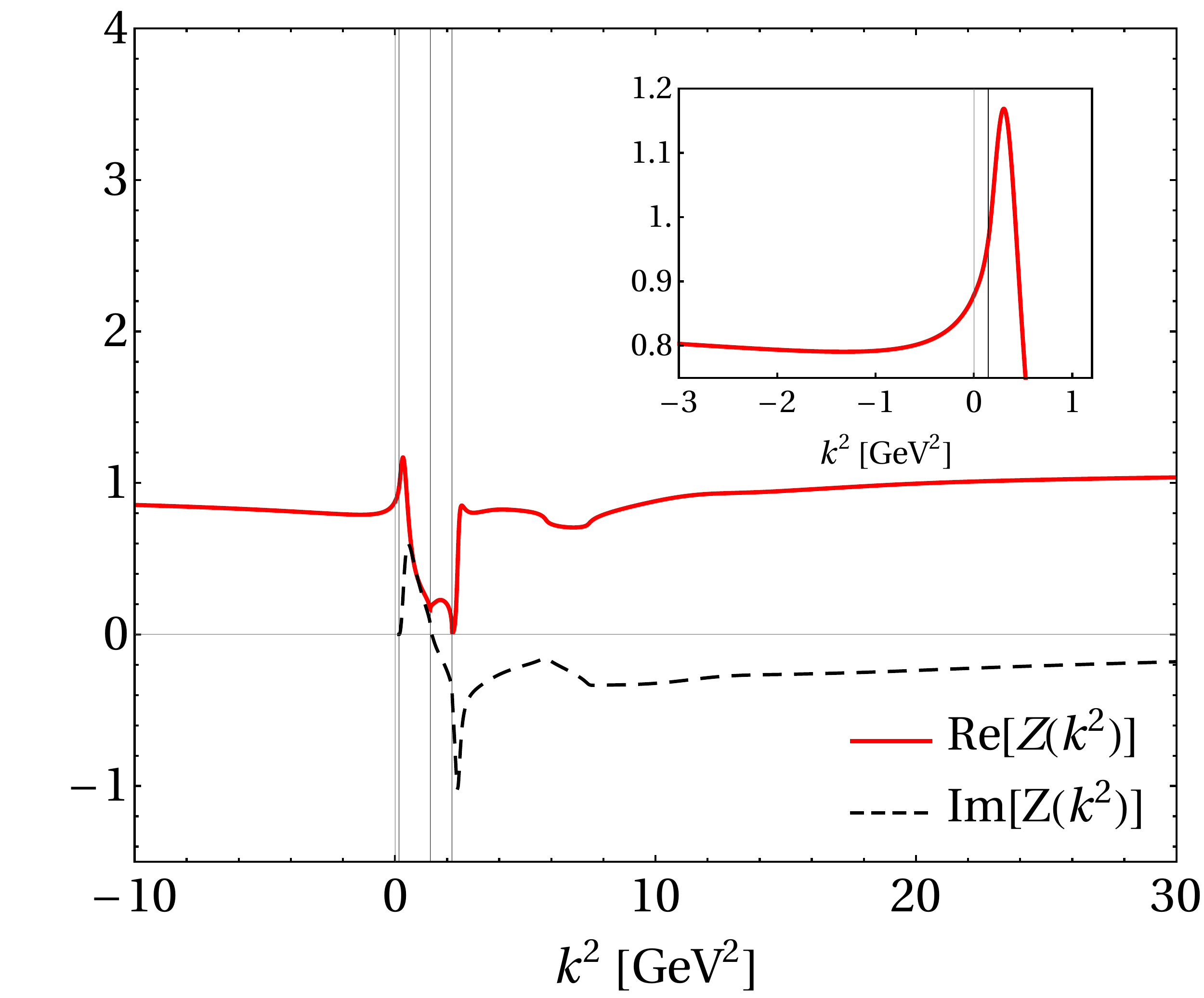}
\par
\caption{Real and imaginary parts of $M(k^2)$ (upper panel) and $Z(k^2)$ (lower panel), for space and time like momenta.  The vertical lines are $\overline m^2_0$, $(\overline m_0+m_g)^2$ and $(\overline m_0+\Lambda)^ 2$ from the thresholds of the driving terms in Eqs.~\eqref{rhoAfull} and \eqref{rhoBfull}.}
\label{MZ}
\end{figure}

We have to add that in Ref.~\cite{Jia:2019kbj} the authors solved the DSE with the rainbow-ladder truncation in Minkowski space, also using Pauli-Villars regularization in the weak coupling regime. In that reference the bare mass $m_B$ was given as the input, while the renormalized mass was evaluated from the self-energy relations. The remaining method was very similar to the approach used here, and it was shown that in the weak coupling regime the structure of the propagator 
is consistent with a simple pole and branch-cuts in the time-like region.

From the Fig.~\ref{MZ} one may see that the imaginary parts of the running mass $M(k^2)$ and wave function renormalization $Z(k^2)$ become nonzero in the time-like momentum region ($k^2>0$), and present discontinuities in their derivatives. The kinks related to these discontinuities occur at the thresholds already discussed for the spectral densities and are indicated by the vertical lines. In these panels it is also shown the results in the space-like region ($k^2<0$). The  inset shows  both curves in the range $0\leq k^2 \leq \overline{m}_0^2$, and the vertical lines indicates the threshold of spectral densities. The source of the rich structure in both $Z(k^2)$ and $M(k^2)$ in the time-like region is associated with the localized strength of the interaction DSE kernel in the IR region below $\sim 1$~GeV, as given by the quark-gluon vertex form factor in Eq.~\eqref{Lambda1}. It is interesting to observe that the Im$[M(k^2)]$ has a pronounced peak for the time-like momentum around the third threshold $\sim 1.5$~GeV. Noteworthy to mention that this value is comparable to the position of the pole at 0.846~GeV of the monopole form fitted to  the quark mass function~\cite{Mello:2017mor}.

The  Phragmén-Lindel\"off theorem~\cite{Tichmarsch}  applies to the Nakanishi integral representation of the self-energies which are bounded, leading to the prediction that the 
asymptotic values for space and time-like regions are identical at high momentum~(see e.g.~\cite{Denig:2012by}), namely $ A(k^2=+\infty)= A(k^2=-\infty)$ and
$ B(k^2=+\infty)= B(k^2=-\infty)$ and they tend to their asymptotic values uniformly. Such properties should be fulfilled by our  results, and indeed we observe in Fig.~\ref{MZ} the mathematical consistency of our numerical calculations, by checking that the  above properties are verified for $M(k^2)$ and $Z(k^2)$  at the large momentum region shown in the figure.

\section{ Final remarks} 
\label{sec:finalremarks}

The phenomena of the mass generation is studied in Minkowski space in a Dyson-Schwinger  model, that incorporates physical scales inspired by lattice QCD results.  The DSE is formulated in the quenched approximation in covariant gauges within the rainbow-ladder truncation with a massive gluon, and a  Pauli-Villars subtraction, which  tunes the infrared physics of the model. The scalar and vector components of the quark propagator and scalar and vector self-energies are described by the  K\"all\'en–Lehman and Nakanishi integral representations, respectively. The associated spectral densities are a solution of a set of inhomogeneous self-consistent equations free of singularities. The essential ingredient to close these equations is the relation between the spectral densities of  the self-energy, propagator,  and the residue at the renormalized mass pole. In the Landau and Feynman gauges the Pauli-Villars term is equivalent to the introduction of a form factor to the quark-gluon vertex, which turns the interpretation of the results more direct. The kernel of the self-consistent equations presents a strong enhancement concentrated in the IR region, mimicking what is   expected from the nonperturbative physics of QCD. 

A detailed numerical study has been performed to explore the dynamical chiral symmetry breaking in the strong coupling constant regime. Solutions were found for the spectral densities in Minkowski space, with the gluon  and  Pauli-Villars masses inspired by Lattice QCD results for the gluon dressing function and the quark-gluon vertex. In addition, the coupling constant and renormalized mass pole, are tuned  to LQCD results for the  quark mass  function in the Landau gauge, concomitantly with a bare mass around 10~MeV. The spectral densities of the scalar and vector  components of the quark propagator show a pronounced negative peak at 0.465~GeV concentrated around a small region between the first threshold at the renormalized mass of 0.38~GeV extending up to $\sim0.6-0.7$~GeV. The positivity  constraints~\cite{Zuber1985} are violated in this region. Furthermore,  the ratio of these scalar and vector densities in this region is found to be the position of the peak, resembling a single mass pole contribution to the propagator. We found that, curiously, the residue at the renormalized mass pole and the analogous integral over the vector density are similar to what was found for a quark propagator model with a mass function described by a monopole form factor fitted to LQCD results~\cite{Mello:2017mor}.

As the quark is not an asymptotic state the spectral densities can in principle violate  the positivity constraints, as we see in this model. The positivity violation in this case is not necessarily equivalent to the confinement, since the latter is a very subtle relation to define in the presence of dynamical quarks~\cite{Dudal:2013vha,Greensite:2011zz}. Therefore, despite being a powerful discriminating tool, this criterion
should be understood as a sufficient, but not necessary condition for the confinement~\cite{Krein:1990sf}.
The wave function renormalization is more critical to be fitted, although our results reflects the IR enhancement of the interaction kernel by the minimum of $Z(k^2)\sim 0.8$ around 1~GeV. 

The spectral densities associated with the self-energies also show a dominant region in infrared, and dominated by the thresholds appearing in the self-consistent equations both in the driving terms as well as in the interaction kernel. These thresholds are defined by the renormalized quark mass pole, gluon and Pauli-Villars masses, which were tuned to approximately reproduce lattice QCD results. We found a curious structure in the scalar densities, namely a ``shark fin'' form reflecting the attempt to build two additional mass poles.
Furthermore, we have explored the analytic structure of the mass function and wave function renormalization in the space and time-like regions. The imaginary part of the mass function seems to suggest a dominant structure with a pronounced peak at time-like momentum of 1.5~GeV, which suggests the possibility that the mass function could have the presence of a mass pole, as the monopole-like form adopted in the QCD-inspired model in~\cite{Mello:2017mor} to fit the LQCD quark mass function.

The present model can be improved in many ways, with a more realistic gluon dressing function and a quark-gluon vertex, which can reproduce LQCD results and at the same time be written in terms of Nakanishi integral representation, for example the analytic forms like the one proposed in~\cite{Li:2019hyv}. 
In the near future we plan to apply the model to the pion taking into account DCSB, by incorporating  dressed constituents with momentum dependent self-energies  in the interaction kernel of the Bethe-Salpeter equation. 
That gives the relevant physical ingredients to further develop the  BS approach for the pion state proposed in~\cite{dePaula:2020qna}, 
to continue the exploration of the Minkowski space structure of this fundamental meson.

\begin{acknowledgments}
The authors thanks Giovanni Salm\`e for the carefull reading of the manuscript,  Pieter Maris and Shaoyang Jia for the  discussions.
This work is a part of the project INCT-FNA proc.~No.~464898/2014-5.
D.C.D. acknowledges the Funda\c{c}\~ao de Amparo \`a Pesquisa do Estado de S\~ao Paulo (FAPESP) grant No.~17/26111-4.
T.F.~acknowledges the CNPq Grant No. 308486/2015-3  and FAPESP under Thematic Projects 2017/05660-0 and  2019/07767-1. 
W.~d.~P.~acknowledges the support from CNPq Grants No. 438562/2018-6 and
No.~313030/2021-9, and  CAPES Grant No.~88881.309870/2018-01.
E.Y.~acknowledges the support from FAPESP Grant No.~2016/25143 and No.~2018/21758-2.
\end{acknowledgments}

\appendix

\section{Spectral densities relations and residue of the propagator}\label{app1}

The connection formulas between $\rho_{A}(\gamma),\rho_{B}(\gamma)$ and $\rho_{v}(\gamma),\rho_{s}(\gamma)$ are obtained by relating the first line of Eq.~(\ref{srenf}) with the definition of $S^ {-1}_f$ given by Eq.~(\ref{Sf1}) with  the self-energies $A(k^2)$ and $B(k^2)$ expressed by their integral representations, and using  the trivial relation $S_{f}^{-1}S_{f} = 1$. One may write:
\begin{align}
S_{v}(\gamma) & =\frac{R}{\gamma-\overline{m}_{0}^{2}+\imath\epsilon}+\int_{0}^{\infty}ds\frac{\rho_{v}(s)}{\gamma-s+\imath\epsilon}=\nonumber \\
 & =\frac{A(\gamma)}{\gamma A^{2}(\gamma)-B^2(\gamma)+\imath\epsilon}\,.\label{sa}
\end{align}
and 
\begin{align}
S_{s}(\gamma) & =\frac{R\,\overline{m}_{0}}{\gamma-\overline{m}_{0}^{2}+\imath\epsilon}+\int_{0}^{\infty}ds\frac{\rho_{s}(s)}{\gamma-s+\imath\epsilon}=\nonumber \\
 & =\frac{B(\gamma)}{\gamma A^{2}(\gamma)-B^2(\gamma)+\imath\epsilon}\,,\label{sb}
\end{align}
which can be rewritten as 
\begin{eqnarray}
S_{v}(\gamma)=\frac{f_{A}(\gamma)-\imath\,\pi\,\rho_{A}(\gamma)}{D(\gamma)}\,,\label{sa1}\nonumber\\
S_{s}(\gamma)=\frac{f_{B}(\gamma)-\imath\,\pi\,\rho_{B}(\gamma)}{D(\gamma)}\,.\label{sb1}
\end{eqnarray}
Here we used the definitions
\begin{align}
D(\gamma & )=\gamma\bigg(f_{A}(\gamma)-\imath\,\pi\,\rho_{A}(\gamma)\bigg)^{2}
\nonumber \\
 & -\bigg(f_{B}(\gamma)-\imath\,\pi\,\rho_{B}(\gamma)\bigg)^{2}\,,
\label{den}
\end{align}
and
\begin{align}
&f_{A}(\gamma)=1+\dashint_{0}^{\infty}ds\frac{\rho_{A}(s)}{\gamma - s}\, ,\label{fainitial}\\
&f_{B}(\gamma)=m_{B}+\dashint_{0}^{\infty}ds\frac{\rho_{B}(s)}{\gamma - s}\, ,\label{fbinitial}
\end{align}
where the functions $f_{A(B)}(\gamma)$ are real and contain principal value integrals.

After  the real and imaginary part of $S_{v}(k^2)$ and $S_{s}(k^2)$ are separated, and performing some manipulations, it is found that:
\begin{align}
 R\delta(\gamma-&\overline{m}_{0}^{2})+\rho_{v}(\gamma)
 =\delta\bigg[\frac{\text{Re}[D(\gamma)]}{f_{A}(\gamma)}\bigg]_{\text{Im}[D(\gamma)]=0}\nonumber \\
 &+\rho_{A}(\gamma)
\,\frac{N_1(\gamma)}{d(\gamma)}  -2\,f_{A}(\gamma)\,\frac{N_2(\gamma)}{d(\gamma)}\, ,
\label{saf}
\end{align}
and 
\begin{align}
R\,\overline{m}_{0}\,\delta(\gamma-&\overline{m}_{0}^{2})+\rho_{s}(\gamma) =\delta\bigg[\frac{\text{Re}[D(\gamma)]}{f_{B}(\gamma)}\bigg]_{\text{Im}[D(\gamma)]=0}\nonumber \\
 & +\rho_{B}(\gamma)\frac{N_1(\gamma)}{d(\gamma)}-2\,f_{B}(\gamma)\frac{N_2(\gamma)}{d(\gamma)}\, ,
\label{sbf}
\end{align}
where
\begin{align}
 &N_1(\gamma)=
 \gamma f_{A}^{2}(\gamma)-\pi^{2}\,\gamma\,\rho_{A}^{2}(\gamma)-f_{B}^{2}(\gamma)+\pi^{2}\,\rho_{B}^{2}(\gamma)\, , \nonumber\\
 &N_2(\gamma)=   \gamma\rho_{A}(\gamma)\,f_{A}(\gamma)-\rho_{B}(\gamma)\,f_{B}(\gamma)\, , \nonumber\\
& d(\gamma)=4\pi^2[N_1(\gamma)]^2+[N_2(\gamma)]^2 \, .
    \end{align}

From Eqs.~\eqref{saf} and \eqref{sbf}
we identify 
\begin{align}
 & \rho_{v}(\gamma)=
 \rho_{A}(\gamma)\frac{N_1(\gamma)}{d(\gamma)}-2\,f_{A}(\gamma)\frac{N_2(\gamma)}{d(\gamma)}\, ,
 \label{rhovv}\\
 & \rho_{s}(\gamma)=\rho_{B}(\gamma) \frac{N_1(\gamma)}{d(\gamma)}-2\,f_{B}(\gamma)
 \frac{N_2(\gamma)}{d(\gamma)}
 \, .
\label{rhoss}
\end{align}
The spectral densities, $\rho_v(\gamma)$ and $\rho_s(\gamma)$, in these equations are different for zero from $\gamma \geq (\overline{m}_0 + \sqrt{\xi}m_{\sigma})^2$, as derived from the support of $\rho_A(\gamma)$ and $\rho_B(\gamma)$.

The residue of the propagator may be evaluated as follows.
Comparing the two sides of (\ref{saf}) and (\ref{sbf}), we find
that: 
\begin{equation}
\gamma-\overline{m}_{0}^{2}=R\frac{\text{Re}[D(\gamma)]}{f_{A}(\gamma)}=R\,\overline{m}_{0}\frac{\text{Re}[D(\gamma)]}{f_{B}(\gamma)}\,,
\end{equation}
for Im$[D(\gamma=\overline{m}_{0}^{2})]=0$. This means that 
\begin{equation}
\overline{m}_{0}^{2}\,\rho_{A}(\overline{m}_{0}^{2})\,f_{A}(\overline{m}_{0}^{2})-\rho_{B}(\overline{m}_{0}^{2})\,f_{B}(\overline{m}_{0}^{2})=0\,,
\end{equation}
which is solved by  $\rho_{A}(\overline{m}_{0}^{2})=\rho_{B}(\overline{m}_{0}^{2})=0$,
or by having $\gamma\geq\overline{m}_{0}^{2}$
such that the spectral densities $\rho_{A}(\gamma)$ and $\rho_{B}(\gamma)$, vanishes
for $\gamma\le(\overline{m}_0 + \sqrt{\xi}m_g)^2$. Then, the argument of the delta appearing in Eqs.~\eqref{saf} and \eqref{sbf} vanishes
if 
$$\overline{m}_{0}^{2}f_{A}^{2}(\overline{m}_{0}^{2})-f_{B}^{2}(\overline{m}_{0}^{2})=0\, .$$
This condition can be achieved once we have
\begin{align}
\overline{m}_{0}f_{A}(\overline{m}_{0}^{2}) & =\overline{m}_{0}+\overline{m}_{0}\dashint_{0}^{\infty}ds\frac{\rho_{A}(s)}{\overline{m}_{0}^{2}-s}\nonumber \\
 & =f_{B}(\overline{m}_{0}^{2})=m_{B}+\dashint_{0}^{\infty}ds\frac{\rho_{B}(s)}{\overline{m}_{0}^{2}-s}\, ,\label{facond}
\end{align}
from where the value of the bare mass $m_B$ may be obtained from Eq.~(\ref{facond}), and it reads:
\begin{equation}
m_B = \overline{m}_0 + \overline{m}_0  \dashint_{0}^{\infty}ds\frac{\rho_{A}(s)}{\overline{m}_{0}^{2}-s} 
-\dashint_{0}^{\infty}ds\frac{\rho_{B}(s)}{\overline{m}_{0}^{2}-s}\, .
\label{mbare}
\end{equation}
Therefore, once the solution of the self-consistent equations, \eqref{rhoAfull} and \eqref{rhoBfull}, are found with the input of the renormalized mass the value of the bare one, $m_B$, is immediately obtained by the above relation.
Further manipulation of Eqs.~\eqref{fainitial} and \eqref{fbinitial}, taking into account the Eq.~\eqref{facond}  to eliminate $m_B$ leads to:
\begin{align}
f_{A}(\gamma) & =1+\dashint_{0}^{\infty}ds\frac{\rho_{A}(s)}{\overline{m}_{0}^{2}-s}\nonumber \\
 & -(\gamma-\overline{m}_{0}^{2})\dashint_{0}^{\infty}ds\frac{\rho_{A}(s)}{(\overline{m}_{0}^{2}-s)(\gamma - s)}\,,
\\
f_{B}(\gamma) & =\overline m_{0}+\overline{m}_{0}\dashint_{0}^{\infty}ds\frac{\rho_{A}(s)}{\overline{m}_{0}^{2}-s}\nonumber \\
 & -(\gamma-\overline{m}_{0}^{2})\dashint_{0}^{\infty}ds\frac{\rho_{B}(s)}{(\overline{m}_{0}^{2}-s)(\gamma - s)}\,.
\end{align}

From the real parts of the denominator $D(\gamma)$ when $\gamma\to \overline{m}_0^2$, and taking into account the previous formulas, one  may finally write
\begin{align}
R^{-1}  =1+&\dashint_{0}^{\infty}\hspace{-0.2cm}ds\frac{\rho_{A}(s)}{\overline{m}_{0}^{2}-s}-2\overline{m}_{0}^{2}\dashint_{0}^{\infty}\hspace{-0.2cm}ds\frac{\rho_{A}(s)}{(\overline{m}_{0}^{2}-s)^{2}}\nonumber \\
 & +2\overline{m}_{0}\dashint_{0}^{\infty}\hspace{-0.2cm}ds\frac{\rho_{B}(s)}{(\overline{m}_{0}^{2}-s)^{2}}\,,
 \label{residue}
\end{align}
which finally close our set of self-consistent integral equations, \eqref{rhoAfull} and \eqref{rhoBfull}, for the spectral densities.


\end{document}